\documentclass[amsmath,amssymb,pre]{revtex4}
\usepackage[]{graphicx} 
\usepackage{latexsym,amsmath,amsfonts,amssymb,bm,empheq}
\usepackage{here}

\begin{document}
\title{Theoretical Basis for Classifying Hyperuniform States of Two-Component Systems}

\author{Hiroshi Frusawa}
\email{frusawa.hiroshi@kochi-tech.ac.jp}
\affiliation{Laboratory of Statistical Physics, Kochi University of Technology, Tosa-Yamada, Kochi 782-8502, Japan.}
\date{\today}
\begin{abstract}
Hyperuniform states of matter exhibit unusual suppression of density fluctuations at large scales, contrasting sharply with typical disordered configurations. Various types of hyperuniformity emerge in multicomponent disordered systems, significantly enhancing their functional properties for advanced applications. This paper focuses on developing a theoretical framework for two-component hyperuniform systems. We provide a robust theoretical basis to identify novel conditions on structure factors for a variety of hyperuniform binary mixtures, classifying them into five distinct types with seven unique states. Our findings also offer valuable guidelines for designing multihyperuniform materials where each component preserves hyperuniformity, added to the overall~hyperuniformity.
\end{abstract}
\maketitle

\section{Introduction}\label{introduction}

Hyperuniform states of matter show an unusual reduction in density fluctuations at large scales, unlike typical disordered systems \cite{rev1,rev2,rev3,frusawa1}. In other words, hyperuniform systems are at a unique critical point where large-scale density fluctuations disappear, which differs from standard thermal critical points where fluctuations become infinitely large \cite{rev1,frusawa1}.
Suppressing large-scale density fluctuations is central to the hyperuniformity concept, which applies to both equilibrium and non-equilibrium phases \cite{rev1,rev2,rev3,frusawa1,two1,two2,two3,two4,two5,two6,two7,two8,poly1,poly2,poly3,ele1,ele2,mul1,mul2,mul3,mul4,mul5,mul6,mul7,mul8,mul9,mul10,non1,non2,non3,non4,non5}.
Since understanding the underlying mechanism that generates hyperuniform patterns is of fundamental interest, hyperuniformity is attracting increasing attention not only in condensed matter physics but also in mathematical physics, including random matrix theory and stochastic geometry theory \cite{rev1,random1,random2,math1,math2,math3}.
Hyperuniform materials have been discovered in crystals, quasicrystals, and novel disordered systems.
This paper focuses on disordered hyperuniform materials, a new class of amorphous matter that exhibits crystal-like behavior by suppressing volume-fraction fluctuations at large scales while remaining isotropic and free of Bragg diffraction peaks.
These materials possess novel photonic, phononic, transport, and mechanical properties, making them useful for various applications \cite{rev1,rev2,rev3}.

Specifically, this paper is concerned with disordered multicomponent systems that preserve various types of hyperuniformity \cite{rev1,two1,two2,two3,two4,two5,two6,two7,two8,poly1,poly2,poly3,ele1,ele2,mul1,mul2,mul3,mul4,mul5,mul6,mul7,mul8,mul9,mul10}.
Phase mixing and separation phenomena are prevalent in multicomponent systems, including alloys, composites, granular media, geological media, biological media, and soft matter such as foams, polymer blends, and colloidal suspensions.
Generalizing the concept of hyperuniformity to two-phase heterogeneous materials, we have observed disordered hyperuniform two-phase materials that exhibit unique transport and electromagnetic properties due to their ability to suppress volume-fraction fluctuations \cite{rev1,rev2,rev3,two1,two2,two3,two4,two5,two6,two7,two8}.
Furthermore, multihyperuniformity emerges in composite systems where multiple distinct subsets are each hyperuniform, in addition to entire point configurations that are hyperuniform \cite{rev1,two1,two2,two3,two4,two5,two6,two7,two8,ele2,mul1,mul2,mul3,mul4,mul5,mul6,mul7,mul8,mul9,mul10}.
Such states have been observed in various natural and synthetic systems, including the photoreceptor patterns in avian retinas \cite{rev1,mul2}.
Multihyperuniformity enhances the functional properties of materials, making them suitable for advanced applications in photonics, electronics, and other fields requiring precise control over material properties \cite{rev1,rev2,rev3}.

Yet, hyperuniformity in multicomponent systems remains elusive.
Here, we restrict ourselves to two-component systems where a variety of hyperuniform states emerge.
We expand the scope to binary solutes within a solvent matrix, which allows us to cover an extensive range of two-component materials.
They include the two-phase systems~\cite{two1,two2,two3,two4,two5,two6,two7,two8}, high-entropy alloys \cite{rev2,rev3,mul10}, binary colloids \cite{rev1,mul3,mul4,mul5,mul6,mul7,mul8}, simple electrolytes \cite{rev1,ele1,ele2}, and polyelectrolytes such as biopolymers \cite{rev1,mul9}. 
Various forms of hyperuniformity have been found in these systems.
While there are models that generate different types of hyperuniformity, hyperuniform states in two-component systems have not been systematically classified~\cite{rev1,two1,two2,two3,two4,two5,two6,two7,two8,ele2,mul1,mul2,mul3,mul4,mul5,mul6,mul7,mul8,mul9,mul10}.
For example, previous studies have demonstrated the relevance of a minimal model that successfully yields multihyperuniformity, one of the various hyperuniform states in multicomponent systems \cite{mul5, mul6}; however, little attention has been paid to a general framework to identify a variety of hyperuniform states in \mbox{mixtures \cite{two1,two2,two3,two4,two5,two6,two7,two8,mul1,mul2,mul3,mul4,mul5,mul6,mul7,mul8,mul9,mul10}}.
A clearer understanding of the type of target hyperuniformity is required to design multicomponent hyperuniform systems.

This paper aims to develop a theoretical framework for two-component hyperuniform systems that addresses the above issues.
We focus on the conditions under which various structure factors (SFs) vanish in the long-wavelength limit where the angular wavenumber $k$ (hereafter simply referred to as the wavenumber) goes to zero. 
To this end, we express the SFs in terms of total correlation functions (TCFs) and direct correlation functions (DCFs) using the Ornstein--Zernike equations for homogeneous systems \cite{liquid}.
Supposing that this theoretical framework applies to quasi-steady heterogeneous states in addition to equilibrium ones, we systematically classify various hyperuniform states in two-component systems based on the theoretical expressions and comprehensively present the possible hyperuniform conditions for each classification.

To this end, this paper is organized as follows.
Section \ref{sec basic} expresses various SFs using DCFs through the Ornstein--Zernike equations as the basic formalism.
Section \ref{sec general} provides general relations necessary for the new hyperuniform conditions to examine unique criteria for two-component hyperuniform systems. Section \ref{sec theoretical} classifies various hyperuniform states in two-component systems based on the developed theory and presents various relations concerning hyperuniform conditions for each group.
In Section \ref{connection}, we compare the obtained relations with previous experimental and simulation results to reorganize the known hyperuniform phenomena and verify the validity of the hyperuniform conditions obtained in Section \ref{sec theoretical}.
Finally, in Section \ref{sec6}, we discuss the remaining challenges and present a summary of the results.

\section{Basic Formalism}\label{sec basic}
Hyperuniformity is characterized by the vanishing of SFs in the limit of $k\rightarrow 0$, or at zero wavenumber.
In Section \ref{subsec ts}, we define the SFs using TCFs.
The Ornstein--Zernike equations in Section \ref{subsec oz} allow us to represent the SFs using DCFs.

\subsection{Total Correlation Functions and Structure Factors}\label{subsec ts}
Let $n_{\mathrm{tot}}$ be the total density in a binary mixture of volume $V$, or the sum of individual densities
$n_{\mathrm{tot}}=n_{\alpha}+n_{\beta}=N/V$,
where $n_{\alpha}$ and $n_{\beta}$ denote the uniform densities of $\alpha$- and $\beta$-components, respectively, and $N$ denotes the total number of particles.
By introducing the number fractions, $\phi_{\alpha}$ and $\phi_{\beta}$, we have
\begin{align}
\label{n l}
n_{l}=\phi_{l}\,n_{\mathrm{tot}}\quad(l=\alpha,\,\beta).
\end{align}
Given the mean particle volume $\overline{w}$ such that $\overline{w}=V/N=n_{\mathrm{tot}}^{-1}$ in two-phase systems, $\phi_l$ ($l=\alpha,\,\beta$) also represents the mean volume fractions 
$\phi_{l}=n_{l}\,\overline{w}$, as seen from Equation~(\ref{n l}). 

The density--density correlation functions are defined using the microscopic density,
$\hat{\rho}\,_{l}(\bm{x})=\sum_{i=1}^N\delta\left(\bm{x}-\bm{r}_i^{l}\right)$ ($l=\alpha,\,\beta$), where $\bm{r}_i^{l}$ denotes the position of the $i$th particle of component $l$.
We calculate the statistical average of the products of microscopic densities over realizations of particle configurations, thereby providing the TCFs \cite{liquid}:
\begin{align}
\left<\,\hat{\rho}\,_{\alpha}(\bm{x})\,\hat{\rho}\,_{\alpha}(\bm{y})\right>
&=n_{\alpha}^2\left\{1+h(r)\right\},
\label{intra alpha}\\
\left<\,\hat{\rho}\,_{\beta}(\bm{x})\,\hat{\rho}\,_{\beta}(\bm{y})\right>
&=n_{\beta}^2\left\{1+h'(r)\right\},
\label{intra beta}\\
\left<\,\hat{\rho}\,_{\alpha}(\bm{x})\,\hat{\rho}\,_{\beta}(\bm{y})\right>
&=n_{\alpha}n_{\beta}\left\{1+\widetilde{h}(r)\right\},
\label{inter}
\end{align}
where $r=|\bm{x}-\bm{y}|$, $h(r)$ and $h'(r)$ will be referred to as the intra-TCFs and $\widetilde{h}(r)$ as the inter-TCF.
We can define the SFs using the TCFs as follows:
the intra- and inter-SFs are given by
\begin{align}
S(k)&=\frac{1}{\phi_{\alpha}\,N}\left<\hat{\rho}_{\alpha}(k)\,\hat{\rho}_{\alpha}(-k)\right>
=1+n_{\alpha}h(k),
\label{def s}\\
S'(k)&=\frac{1}{\phi_{\beta}\,N}\left<\hat{\rho}_{\beta}(k)\,\hat{\rho}_{\beta}(-k)\right>
=1+n_{\beta}h'(k),
\label{def s'}\\
\widetilde{S}(k)&=\frac{1}{\sqrt{\phi_{\alpha}\,\phi_{\beta}}\,N}\left<\hat{\rho}_{\alpha}(k)\,\hat{\rho}_{\beta}(-k)\right>=\sqrt{n_{\alpha}n_{\beta}}\,\widetilde{h}(k),
\label{def tilde s}
\end{align}
respectively.

We also introduce either the sum or the difference of microscopic densities,
\begin{align}
\hat{\rho}\,_{\sigma}(k)=\hat{\rho}_{\alpha}(k)+\sigma\hat{\rho}_{\beta}(k)\quad(\sigma=+,\,-),
\label{def rho sigma}
\end{align}
with which the total SFs are defined for $\sigma=+$ and $-$ as follows:
\begin{align}
S_{\sigma}^{\mathrm{tot}}(k)&=\frac{1}{N}\left<\hat{\rho}\,_{\sigma}(k)\,\hat{\rho}\,_{\sigma}(-k)\right>,
\label{def s total}
\end{align}
providing
\begin{align}
S_{\sigma}^{\mathrm{tot}}(k)&=\phi_{\alpha}\,S(k)+\phi_{\beta}\,S'(k)+2\sigma\sqrt{\phi_{\alpha}\,\phi_{\beta}}\,\widetilde{S}(k),
\label{rep s total}
\end{align}
due to Equations (\ref{def s})--(\ref{def tilde s}).
It should also be noted that
\begin{align}
S_{-\sigma}^{\mathrm{tot}}(k)=S_{\sigma}^{\mathrm{tot}}(k)-4\sigma\sqrt{\phi_{\alpha}\,\phi_{\beta}}\,\widetilde{S}(k) 
\label{negative total}
\end{align}
for later use.

\subsection{The Ornstein--Zernike Equations and Their Solutions}\label{subsec oz}
We introduce DCFs that directly contribute to the TCFs:
while $c(r)$ and $c'(r)$ denote the intra-DCFs, $\widetilde{c}(r)$ denotes the inter-DCF.
The Ornstein--Zernike equations for two-component systems relate the Fourier transforms of the DCFs to those of the TCFs as follows \cite{liquid}:
\begin{flalign}
\label{oz1}
h(k)&=c(k)+n_{\mathrm{tot}}\,\left\{
\phi_{\alpha}\,c(k)h(k)+\phi_{\beta}\,\widetilde{c}(k)\widetilde{h}(k)\right\},
\\
\label{oz2}
h'(k)&=c'(k)+n_{\mathrm{tot}}\,\left\{
\phi_{\beta}\,c'(k)h'(k)+\phi_{\alpha}\,\widetilde{c}(k)\widetilde{h}(k)\right\},
\\
\label{oz3}
\widetilde{h}(k)&=\widetilde{c}(k)+n_{\mathrm{tot}}\,\left\{
\phi_{\alpha}\,c(k)\widetilde{h}(k)+\phi_{\beta}\,\widetilde{c}(k)h'(k)\right\},\\
\label{oz4}
&=\widetilde{c}(k)+n_{\mathrm{tot}}\,\left\{
\phi_{\beta}\,c'(k)\widetilde{h}(k)+\phi_{\alpha}\,\widetilde{c}(k)h(k)\right\}.
\end{flalign}
The Ornstein--Zernike equations allow us to express the TCFs using the DCFs, therefore yielding the SFs defined by Equations (\ref{def s})--(\ref{def tilde s}) (see Appendix \ref{appendix} for details of the derivation):
\begin{align}
\label{s by c}
&S(k)=1+n_{\alpha}h(k)=\frac{\mathcal{C}_{\beta}(k)}{D(k)},\\
\label{s' by c}
&S'(k)=1+n_{\beta}h'(k)=\frac{\mathcal{C}_{\alpha}(k)}{D(k)},\\
\label{tilde s by c}
&\widetilde{S}(k)=\sqrt{n_{\alpha}n_{\beta}}\,\widetilde{h}(k)=\frac{\sqrt{n_{\alpha}n_{\beta}}\,\widetilde{c}(k)}{D(k)},\\
\label{total s by c}
&S_{\sigma}^{\mathrm{tot}}(k)=\frac{\phi_{\alpha}\,\mathcal{C}_{\beta}(k)+\phi_{\beta}\,\mathcal{C}_{\alpha}(k)+2\sigma\sqrt{\phi_{\alpha}\,\phi_{\beta}}\,\widetilde{c}(k)}{D(k)},
\end{align}
where
\begin{align}
\mathcal{C}_{\alpha}(k)&=1-n_{\alpha}c(k),
\label{def c alpha}\\
\mathcal{C}_{\beta}(k)&=1-n_{\beta}c'(k),
\label{def c beta}\\
D(k)&=\mathcal{C}_{\alpha}(k)\,\mathcal{C}_{\beta}(k)-n_{\alpha}n_{\beta}\widetilde{c}(k)^2.
\label{def D}
\end{align}
It is significant to note that $D(k)$ is an indicator of stability for uniformity in two-component systems:
homogeneous mixtures satisfy $D(k)>0$;
otherwise, divergent density fluctuations are observed at $D(k)=0$, as seen from Equations (\ref{s by c})--(\ref{tilde s by c}), and the uniform systems become linearly unstable when $D(k)<0$ \cite{liquid2}.

In what follows, we consider stationary SFs \cite{frusawa stability1,frusawa stability2} that properly provide time-invariant thermodynamic quantities, including isothermal compressibility $\chi_T$, regardless of $D(k)$.
As typical indicators of thermodynamic stability, this paper adopts not only $\chi_T$ but also the mean-square fluctuation $\sigma_N^2$ in the total number of particles. 
We will assess the thermodynamic stability of hyperuniform binary mixtures using the following relations~\cite{mul7,comp1,comp2,comp3}:
\begin{align}
\frac{\sigma_N^2}{N}&=\lim_{k\rightarrow 0}S_+^{\mathrm{tot}}(k)\geq 0,
\label{def variance}\\
n_{\mathrm{tot}}k_BT\chi_T&=\lim_{k\rightarrow 0}\frac{1}{\phi_{\alpha}\mathcal{C}_{\beta}(k)+\phi_{\beta}\mathcal{C}_{\alpha}(k)-2\sqrt{\phi_{\alpha}\,\phi_{\beta}}\,\widetilde{c}(k)}\geq 0,
\label{def chi}
\end{align}
where $k_BT$ denotes the thermal energy.

\section{General Relations: When SFs Vanish with {\itshape D}({\itshape k}) of Finite Extent}\label{sec general}
Equations (\ref{s by c})--(\ref{total s by c}) imply that SFs vanish in the limit of $D(k)\rightarrow\infty$; however, this section deals with the disappearances of SFs while keeping $D(k)$ finite.
We will present valuable relations that generally hold when some SFs vanish.  

\subsection{Partial Disappearance of Total SFs with Inter-Correlations}\label{subsec partial}
Given the expression (\ref{rep s total}) of a total SF, $S_{\sigma}^{\mathrm{tot}}(k)$ ($\sigma=+,\,-$), we have the following relation at $k_0$, where $S_{\sigma}^{\mathrm{tot}}(k)$ disappears without the divergence of $D(k)$:
\begin{align}
\frac{-2\sigma\sqrt{\phi_{\alpha}\phi_{\beta}}\,\widetilde{S}(k_0)}{\phi_{\alpha}S(k_0)+\phi_{\beta}S'(k_0)}
=\frac{-2\sigma\sqrt{\phi_{\alpha}\phi_{\beta}}\sqrt{n_{\alpha}n_{\beta}}\,\widetilde{c}(k_0)}{\phi_{\alpha}\mathcal{C}_{\beta}(k)+\phi_{\beta}\mathcal{C}_{\alpha}(k)}=1,
\label{zero read}
\end{align}
where use is made of Equations (\ref{s by c})--(\ref{def c beta}).

Equation (\ref{zero read}) implies that
\begin{align}  
\sigma\,\widetilde{S}(k_0)\leq 0 
\label{ineq1}
\end{align}
because of $S(k)\geq 0$ and $S'(k)\geq 0$ by definition.
Since the other total SF $S_{-\sigma}^{\mathrm{tot}}(k)$ given by Equation (\ref{rep s total}) reduces to
\begin{align}
S_{-\sigma}^{\mathrm{tot}}(k_0)
=-4\sigma\sqrt{\phi_{\alpha}\,\phi_{\beta}}\,\widetilde{S}(k_0),
\label{rep s -total}
\end{align}
Equation (\ref{ineq1}) leads to
\begin{align}
S_{-\sigma}^{\mathrm{tot}}(k_0)\geq 0,
\label{ineq2}
\end{align}
stating that $S_{-\sigma}^{\mathrm{tot}}(k_0)>0$ is necessarily valid as long as there exist inter-correlations: \linebreak$\widetilde{S}(k_0)\neq 0$. 

\subsection{Stability in Terms of {\itshape D}({\itshape k})}\label{subsec stability}
There is a restriction on the sign of $D(k)$ not only at $k_0$ (see Section \ref{subsec partial}) but also at $k_1$, where either or both of the intra-SFs ($S(k)$ and $S'(k)$) are set to zero with $D(k)$ of finite extent.
As seen below, two-component systems never enter the stable region of $D(k)>0$ at $k_0$ and~$k_1$.

In the absence of divergence in $D(k)$, Equations (\ref{s by c}) and (\ref{s' by c}) imply that $S(k_1)=0$ and $S'(k_1)=0$ read
\begin{align}
\mathcal{C}_{\beta}(k_1)&=0\label{partial zero1}\\    
\mathcal{C}_{\alpha}(k_1)&=0,\label{partial zero2}
\end{align}
respectively.
It is readily seen that
\begin{equation}
D(k_1)=-n_{\alpha}n_{\beta}\widetilde{c}(k_1)^2\leq 0
\label{ineq3}
\end{equation}
when either either or both of Equations (\ref{partial zero1}) and (\ref{partial zero2}) are satisfied.

Meanwhile, the total SF vanishes at $k_0$, where Equation (\ref{zero read}) is rewritten to
\begin{align}
\phi_{\alpha}\mathcal{C}'_{\sigma}(k_0)+\phi_{\beta}\mathcal{C}_{\sigma}(k_0)&=2\phi_{\alpha}\mathcal{C}'_{\sigma}(k_0)+\Delta\mathcal{C}(k_0)\nonumber\\
&=2\phi_{\beta}\mathcal{C}_{\sigma}(k_0)-\Delta\mathcal{C}(k_0)\nonumber\\
&=0,
\label{zero read3}
\end{align}
using
\begin{align}
&\mathcal{C}_{\sigma}(k)=\mathcal{C}_{\alpha}(k)+\sigma n_{\alpha}\widetilde{c}(k),
\label{def r}\\
&\mathcal{C}'_{\sigma}(k)=\mathcal{C}_{\beta}(k)+\sigma n_{\beta}\widetilde{c}(k),
\label{def r'}\\
&\Delta\mathcal{C}(k)
=\phi_{\beta}-\phi_{\alpha}+\phi_{\alpha}\phi_{\beta}\,n_{\mathrm{tot}}\left\{c'(k)-c(k)\right\}.
\label{def delta r}
\end{align}
In obtaining the above expressions of $\mathcal{C}_{\sigma}(k)$, $\mathcal{C}'_{\sigma}(k)$, and $\Delta\mathcal{C}(k)$, use was made of Equation (\ref{n l}) and the following relation:
\begin{align}
\sqrt{\phi_{\alpha}\phi_{\beta}}\,\sqrt{n_{\alpha}n_{\beta}} 
=\phi_{\alpha}\phi_{\beta}\,n_{\mathrm{tot}}=\phi_{\alpha}\,n_{\beta}=\phi_{\beta}\,n_{\alpha}.
\label{n cross}
\end{align}
Rearrangement of Equations (\ref{zero read3})--(\ref{def r'}) yields
\begin{align}
\mathcal{C}_{\alpha}(k_0)&=-\sigma n_{\alpha}\,\widetilde{c}(k_0)+\frac{\Delta\mathcal{C}(k_0)}{2\phi_{\beta}},
\label{pre d1}\\
\mathcal{C}_{\beta}(k_0)&=-\sigma n_{\beta}\,\widetilde{c}(k_0)-\frac{\Delta\mathcal{C}(k_0)}{2\phi_{\alpha}}.
\label{pre d2}
\end{align}
Inserting Equations (\ref{pre d1}) and (\ref{pre d2}) into Equation (\ref{def D}) at $k_0$, we obtain
\begin{align}
D(k_0)=-\frac{\Delta\mathcal{C}(k_0)^2}{4\phi_{\alpha}\phi_{\beta}}\leq 0.
\label{d k0}
\end{align}
For $c(k)=c'(k)$ in Equation (\ref{def delta r}), Equation (\ref{d k0}) reduces to
\begin{align}
D(k_0)=-\frac{(1-2\phi_{\alpha})^2}{4\phi_{\alpha}(1-\phi_{\alpha})}.
\label{d k0 redu}
\end{align}

\begin{figure}[H]
\begin{center}
\includegraphics[
width=16cm
]{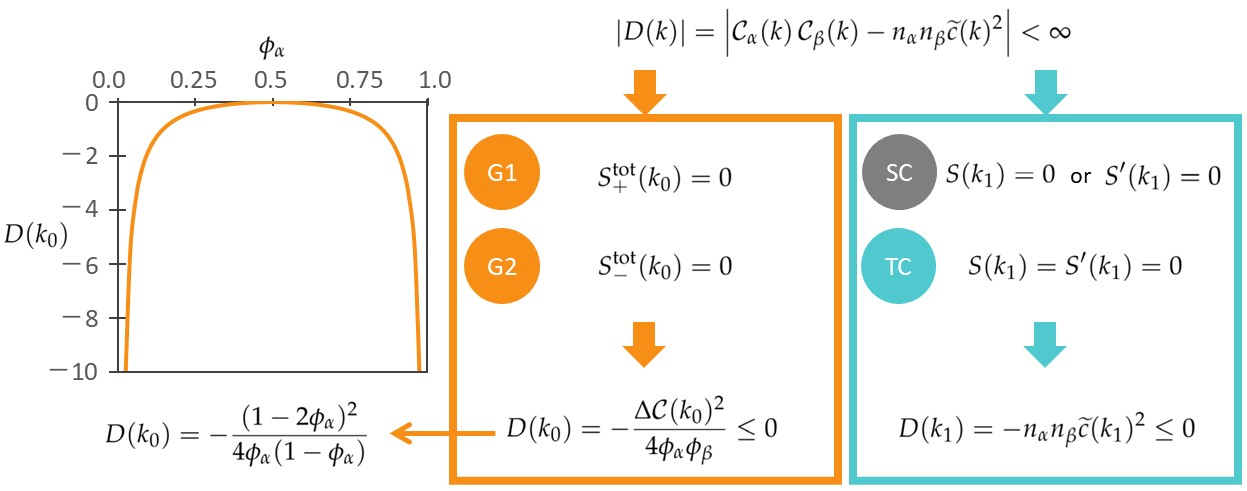}
\end{center}
\caption{
Illustrations
of the two cases in Section \ref{subsec stability} where some SFs disappear at $k_0$ or $k_1$ with $D(k)$ of finite extent: $|D(k)|<\infty$ (see the main text or the nomenclature for the abbreviations of G1, G2, SC, and TC).
}
\label{fig stability}
\end{figure}

Figure \ref{fig stability} illustrates two cases where hyperuniform states emerge without divergence of $D(k)$ while depicting the profile of $D(k_0)$ given by Equation (\ref{d k0 redu}).
Figure \ref{fig stability} summarizes that our general formulations verified $D(k)\leq 0$ at $k_0$ and $k_1$, where the SFs equal zero due to the numerators instead of the denominators $D(k)$ in Equations (\ref{s by c})--(\ref{total s by c}).
The non-positivity conditions on $D(k)$ reveal that uniform two-component systems are marginal or unstable at $k_0$ and $k_1$.
The encircled abbreviations, `SC' and `TC', denote single-component and two-component hyperuniformities, respectively, at $k_0\rightarrow 0$.
Meanwhile, `G1' and `G2' signify global hyperuniformities for $\sigma=+$ (the sum of component densities) and $\sigma=-$ (the density difference), respectively, at $k_1\rightarrow 0$: $\lim_{k\rightarrow 0}S_{\sigma}^{\mathrm{tot}}=0$.
The orange curve depicts the relation (\ref{d k0 redu}) between $D(k_0)$ and $\phi_{\alpha}$. 

\section{Theoretical Predictions of Hyperuniform States in Two-Component~Systems}\label{sec theoretical}
\subsection{Tentative Classification of Two-Component Hyperuniformity}
\begin{figure}[H]
\begin{center}
\includegraphics[
width=14cm
]{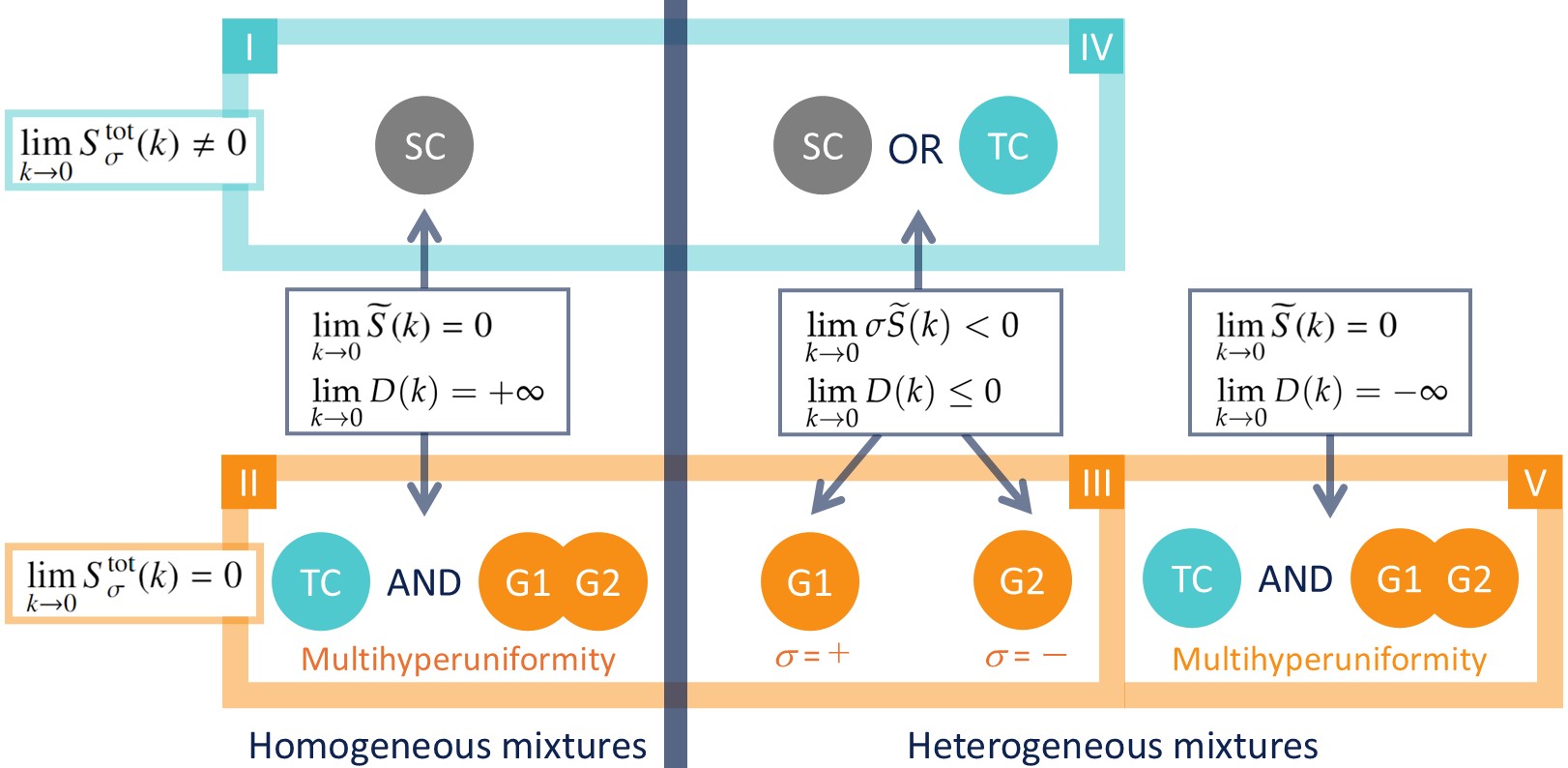}
\end{center}
\caption{Tentative classification flow into various hyperuniform states.
}
\label{fig summary}
\end{figure}

The general results in Section \ref{sec general} become available for hyperuniform states in two-component systems at $k_0,\,k_1\rightarrow 0$.
This section provides theoretical predictions about the properties of various hyperuniform states when $D(k)$ remains finite in the zero-wavenumber limit in addition to going to infinity.
As a summary of Sections \ref{subsec com}--\ref{subsec type5}, Figure~\ref{fig summary} lists possible hyperuniform systems classified into four types of six states distinguished by combinations of single-component, two-component, and global hyperuniformity, prior to assessing the thermodynamic stability in Section \ref{subsec therm}.

In Figure \ref{fig summary}, the upper and lower states are colored blue and orange, respectively, to distinguish between them based on whether global hyperuniformity is preserved.
The vertical thick line also separates the hyperuniform states into two groups depending on the sign of stability indicator $D(k)$ in the zero-wavenumber limit.
The left group satisfies the relation $\lim_{k\rightarrow 0}S_{\sigma}^{\mathrm{tot}}=0$ with $D(0)>0$, thereby ensuring homogeneity. 
The right group, on the other hand, considers mixtures in the region of $\lim_{k\rightarrow 0}D(k)\leq 0$, where homogeneous systems are unstable and tend to become heterogeneous, regardless of whether the inter-SF $\widetilde{S}(k)$ goes to zero in the zero-wavenumber limit or not.
In Sections \ref{subsec type12}--\ref{subsec type5}, we will confirm the validity of relations for $\widetilde{S}(k)$ and $D(k)$ given in the tentative classification flow of Figure \ref{fig summary}.

The seven hyperuniform states are classified into five types as follows:
\begin{itemize}
\item Type I: Single-component hyperuniformity appears without inter-correlations.
\item Type II: There emerge multihyperuniform states where not only two-component hyperuniformity but also both global hyperuniformities denoted by `G1' and `G2' are preserved in the absence of inter-correlations (i.e., $\widetilde{S}(0)= 0$).
\item Type III: There are two kinds of globally hyperuniform states that vary depending on the sign of $\widetilde{S}(0)$.
\item Type IV: Either single-component or two-component hyperuniformity is observed, though global hyperuniformity is lost due to the presence of inter-correlations (i.e., $\widetilde{S}(0)\neq 0$).
\item Type V: This type preserves multihyperuniformity when taking the opposite limit of $\lim_{k\rightarrow0}D(k)=-\infty$ to that of Type-II multihyperuniformity.
\end{itemize}

\subsection{Compatibility of Global and Two-Component Hyperuniformity}\label{subsec com}
There are a variety of hyperuniform states in two-component systems.
The relations at $k_0$ in Sections \ref{subsec partial} and \ref{subsec stability} apply to two types of global hyperuniformity fulfillment,
\begin{align}
\lim_{k_0\rightarrow 0}S_{\sigma}^{\mathrm{tot}}(k_0)=0
\label{global1},
\end{align}
for $\sigma=+$ or $-$. Meanwhile, the relations at $k_1$ in Section \ref{subsec stability} apply to single-component or two-component hyperuniformity that is defined by either or both of the\linebreak following equations:
\begin{align}
\lim_{k_1\rightarrow 0}S(k_1)=0,
\label{multi1}\\
\lim_{k_1\rightarrow 0}S'(k_1)=0.
\label{multi2}
\end{align}
It is necessary for preserving global and two-component hyperuniformity (i.e., multihyperuniformity) to meet the requirements in Equations (\ref{global1})--(\ref{multi2}) simultaneously.
Given the expression (\ref{rep s total}) of $S_{\sigma}^{\mathrm{tot}}(k)$, it also turns out that the inter-SF $\widetilde{S}(k)$ should disappear in the zero-wavenumber limit in order to form multihyperuniform states in\linebreak two-component~systems.

Here, it should be remembered that the spectral densities for phase volumes of two-phase systems can be obtained by multiplying the corresponding SFs by the factor $\phi_l\phi_m/\sqrt{n_ln_m}$ ($l,\,m=\alpha,\,\beta$) in the zero-wavenumber limit, where the appropriate subscripts, $l$ and $m$, are chosen according to the subscripts on the left-hand sides of \mbox{Equations (\ref{intra alpha})--(\ref{inter}) \cite{rev1,two1,two2,two3,two4,two5,two6,two7,two8}}.
From the simple proportionality between the SFs and spectral densities, we surmise that the theoretical consequences due to the SF-based discussions serve to predict features of hyperuniform states in two-phase systems characterized by scalar fields and their spectral densities.
In fact, the absence of inter-correlations in multihyperuniform states has been concluded from the previous theoretical discussions regarding spectral densities in two-phase systems \cite{rev1,two1,two2,two3,two4,two5,two6,two7,two8}.

\begin{figure}[H]
\begin{center}
\includegraphics[
width=12cm
]{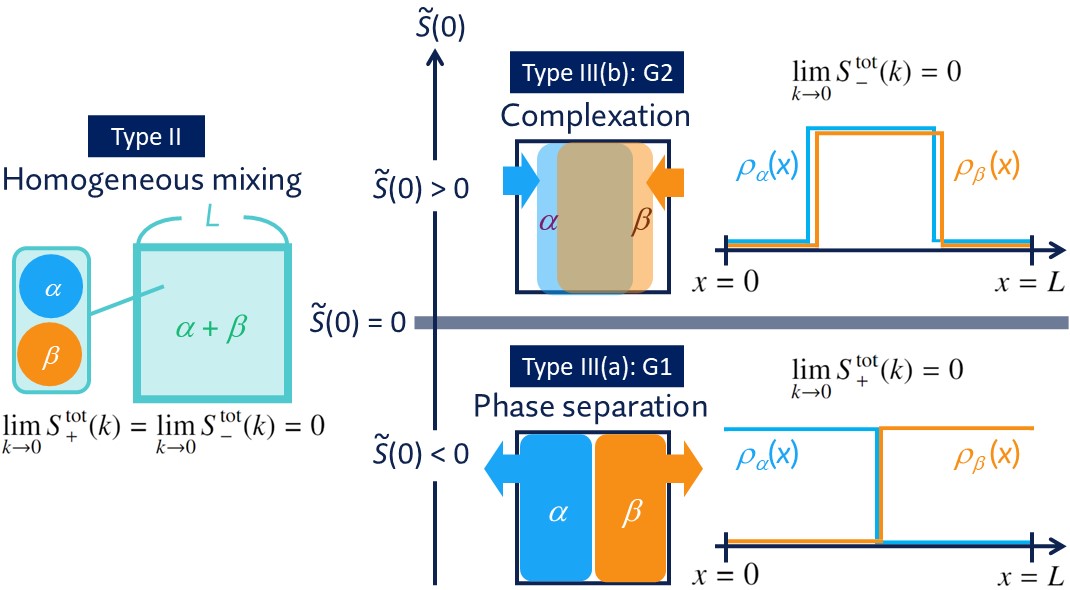}
\end{center}
\caption{A schematic of three states with global hyperuniformity.
}
\label{fig heterogeneity}
\end{figure}

On the contrary, in the presence of inter-correlations, only either global or two-component hyperuniformity emerges, necessarily accompanied by the following heterogeneity:
complex aggregates are formed for $\widetilde{S}(k)>0$, whereas phase separation occurs for $\widetilde{S}(k)<0$.
Figure \ref{fig heterogeneity} depicts the density distributions in globally hyperuniform systems (Types II and III in Figure \ref{fig summary}) to illustrate the difference between three states that satisfy $\widetilde{S}(0)=0$, $\widetilde{S}(0)>0$, and $\widetilde{S}(0)<0$, with global hyperuniformity preserved.

In Figure \ref{fig heterogeneity}, we use schematics of Type-II and -III systems with vertical and horizontal axes.
While the vertical axis represents the sign of inter-SF $\widetilde{S}(0)$ in the zero-wavenumber limit, the horizontal axis depicts an inhomogeneous density distribution of each component on a macroscopic scale.
Homogeneous mixing of $\alpha$- and $\beta$-components occurs at $\widetilde{S}(0)=0$, where no macroscopic correlations exist between densities of different species.
Therefore, multihyperuniform systems emerge in homogeneous mixtures (Type-II states).
On the other hand, global hyperuniformity with non-zero inter-SF ($\widetilde{S}(0)\neq 0$) is referred to as the Type-III state in Figure \ref{fig summary}. 
Since the vertical axis denotes the change from a negative to a positive value of $\widetilde{S}(0)$, we can see two kinds of Type-III hyperuniformities, or the global hyperuniformities for $\sigma=+$ (Type III(a): G1) and $\sigma=-$ (Type III(b): G2), in the lower and upper parts along the vertical axis, respectively.
The horizontal $x$-axis representing the whole system allows us to depict two types of density distributions of $\alpha$- and $\beta$-components schematically, considering that $\widetilde{S}(0)<0$ and $\widetilde{S}(0)>0$ indicate the occurrences of macroscopic phase separation and complexation, respectively.
The density profiles at $\phi_{\alpha}=\phi_{\beta}=0.5$ illustrate that the density sum ($\rho_{\alpha}(x)+\rho_{\beta}(x)$) is more likely to be uniform than the density difference ($\rho_{\alpha}(x)-\rho_{\beta}(x)$) for Type III(a) and vice versa for Type~III(b).

\subsection{Types I and II: Stability Indicator {\itshape D}({\itshape k}) Divergent in Absence of Inter-Correlations}\label{subsec type12}
We have a trivial solution to the Ornstein--Zernike Equations (\ref{oz3}) and (\ref{oz4}): \linebreak\mbox{$\widetilde{h}(k)=\widetilde{c}(k)=0$}.
Equation (\ref{tilde s by c}) also indicates that, other than the trivial result of $\lim_{k\rightarrow 0}\widetilde{c}(k)=0$, the divergence of the stability indicator meets the condition for the disappearance of inter-SF in the zero-wavenumber limit:
\begin{align}
\lim_{k\rightarrow 0}D(k)=+\infty
\label{divergence D}
\end{align}
is a sufficient condition for
\begin{align}
\lim_{k\rightarrow 0}\widetilde{S}(k)=0.
\label{cross zero}
\end{align}
To see this more clearly, we rewrite Equation (\ref{tilde s by c}) to 
\begin{align}
\widetilde{S}(k)=\frac{\sqrt{n_{\alpha}n_{\beta}}\,\widetilde{c}(k)}{\mathcal{C}_{\alpha}(k)\,\mathcal{C}_{\beta}(k)}\left\{
1-\frac{n_{\alpha}n_{\beta}\widetilde{c}(k)^2}{\mathcal{C}_{\alpha}(k)\,\mathcal{C}_{\beta}(k)}
\right\}^{-1},
\label{tilde s2}
\end{align}
making it obvious that Equation (\ref{cross zero}) is satisfied when
\begin{align}
\lim_{k\rightarrow 0}\mathcal{C}_{\alpha}(k)\,\mathcal{C}_{\beta}(k)=+\infty
\label{divergence cc}
\end{align}
with $\widetilde{c}(k)$ being finite, or
\begin{align}
\lim_{k\rightarrow 0}\frac{n_{\alpha}n_{\beta}\widetilde{c}(k)^2}{\mathcal{C}_{\alpha}(k)\,\mathcal{C}_{\beta}(k)}=0.
\label{upper1}
\end{align}
The expression (\ref{def D}) of $D(k)$ under the condition of Equations (\ref{divergence cc}) and (\ref{upper1}) allows us to verify Equation (\ref{divergence D}) as well as the positivity of $D(k)$.
We have thus validated the above relation between Equations (\ref{divergence D}) and (\ref{cross zero}):
it has been proven that the inter-SF in the zero-wavenumber limit, $\widetilde{S}(0)$, goes to zero in the limit of Equation (\ref{divergence cc}).

There are two cases for reaching the limit of Equation (\ref{divergence cc}): 
\begin{itemize}
\item[(i)]  Either $\widetilde{C}_{\alpha}(0)$ or $\widetilde{C}_{\beta}(0)$ contributes to the relation (\ref{divergence cc}), whereas the other non-divergent factor $\widetilde{C}_{l}(0)$ ($l=\alpha$ or $\beta$) is in the order of $\sqrt{n_{\alpha}n_{\beta}}\,\widetilde{c}(0)$. 
\item[(ii)] Both factors of $\widetilde{C}_{\alpha}(0)$ and $\widetilde{C}_{\beta}(0)$ are extremely large: $\widetilde{C}_{\alpha}(0),\,\widetilde{C}_{\beta}(0)\gg1$.
\end{itemize}
These two cases of limiting behaviors for the DCFs create the following distinct features:
while single-component hyperuniformity emerges in Type-I systems, multihyperuniformity is preserved in Type-II systems.

(Type I)
For example, let $\mathcal{C}_{\alpha}\gg 1$ and $\mathcal{C}_{\beta}(0)\sim\sqrt{n_{\alpha}n_{\beta}}\,\widetilde{c}(0)$ be taken as a condition of the case (i).
The intra-SF $S(k)$ of the $\alpha$-component is approximated by 
\begin{align}
\lim_{k\rightarrow 0}S'(k)
&=\lim_{k\rightarrow 0}\frac{\mathcal{C}_{\alpha}(k)}{\mathcal{C}_{\alpha}(k)\,\mathcal{C}_{\beta}(k)}\left\{
1-\frac{n_{\alpha}n_{\beta}\widetilde{c}(k)^2}{\mathcal{C}_{\alpha}(k)\,\mathcal{C}_{\beta}(k)}
\right\}^{-1}\nonumber\\
&\approx\frac{1}{\mathcal{C}_{\beta}(k)}\neq 0:
\label{s'2}
\end{align}
$S'(k)$ does not vanish while Equations (\ref{multi1}) and (\ref{cross zero}) are valid.
Thus, we have demonstrated that Type-I systems preserve single-component hyperuniformity in the sense that neither global nor two-component hyperuniformity is observed due to Equation (\ref{s'2}).

(Type II)
It is easily confirmed that Equations (\ref{multi1}) and (\ref{multi2}) are valid in Type II, where $1/\mathcal{C}_l(0)\rightarrow 0$ ($l=\alpha$, $\beta$).
Combining Equations (\ref{multi1}), (\ref{multi2}), and (\ref{cross zero}), it turns out that Type-II states are multihyperuniform.
In addition, the definitions in Equations (\ref{def c alpha}) and (\ref{def c beta}) verify that $\mathcal{C}_{\alpha}(0)\,\mathcal{C}_{\beta}(0)$ can be approximated as $n_{\alpha}n_{\beta}\,c(0)c'(0)$ in the case (ii).
As a consequence, the relation (\ref{upper1}) reads
\begin{align}
\lim_{k\rightarrow 0}\left\{c(k)c'(k)-\widetilde{c}(k)^2\right\}>0.
\label{multi condition}
\end{align}
Replacing the DCFs in Equation (\ref{multi condition}) with the bare interaction potentials in the spirit of the random phase approximation, Equation (\ref{multi condition}) is consistent with the previous condition for multihyperuniformity \cite{mul6}.

\subsection{Type III: Global Hyperuniformity with Inter-Correlations}\label{subsec type3}
When considering $D(0)$ of finite extent, we should note that Equation (\ref{global1}) is transformed into Equation (\ref{zero read}) at $k_0=0$.
We also see from Equation (\ref{tilde s by c}) that the inter-SF $\widetilde{S}(k)$ does not vanish, with $D(0)$ being finite unless $\widetilde{c}(k)= 0$.
Thus, Equation (\ref{zero read}) at $k_0=0$ represents the global hyperuniformity with inter-correlations, and the associated \mbox{relations (\ref{ineq1}) and (\ref{ineq2})} read
\begin{align}
\label{result ineq1}
&\sigma\,\widetilde{S}(0)<0,\\
\label{result ineq2}
&S_{-\sigma}^{\mathrm{tot}}(0)>0,
\end{align}
respectively, in globally hyperuniform systems with inter-correlations, where Equation (\ref{global1}) is valid despite $\widetilde{S}(0)\neq 0$.

Specifically, the relations (\ref{result ineq1}) and (\ref{result ineq2}) characterize the global hyperuniformity with inter-correlations as follows (see also Figure \ref{fig summary}):
while the first kind of global hyperuniformity focuses on the density distribution regarding the sum of each component's density (Type III(a): G1, or $\sigma=+$ in Equation (\ref{def rho sigma})), the other kind of global hyperuniformity treats that of the difference between each component's density (Type III(b): G2, or $\sigma=-$ in Equation (\ref{def rho sigma})).
Figure \ref{fig heterogeneity} shows that the species of global Type-III hyperuniformity varies depending on whether different components are likely to be excluded from each other (Type III(a): G1, or $\widetilde{S}(0)<0$) or to be aggregated to form complexes (Type III(b): G2, or $\widetilde{S}(0)>0$) on a macroscopic scale.
Namely, the globally hyperuniform systems with inter-correlations have the tendency to spatial heterogeneity.

The preference for inhomogeneous density distribution reflects the stability indicator $D(k)$ that is restricted to
\begin{align}
\lim_{k_0\rightarrow 0}D(k_0)\leq 0,
\label{result ineq3}
\end{align}
due to the expression (\ref{d k0}).
Remarkably, the divergence of SFs is predicted in symmetric two-component systems:
it follows from Equations (\ref{tilde s by c}), (\ref{rep s -total}) and (\ref{d k0}) that
\begin{align}
\lim_{\Delta\mathcal{C}\rightarrow 0}\sigma\widetilde{S}(0)=-\infty,
\label{inf1}\\
\lim_{\Delta\mathcal{C}\rightarrow 0}S_{-\sigma}^{\mathrm{tot}}(0)=+\infty,
\label{inf2}
\end{align}
due to $D(0)=0$, similar to the previous results of two-replica systems \cite{frusawa2}.
The relations~(\ref{result ineq3}) to (\ref{inf2}) reveal that uniform two-component systems are inevitably marginal or unstable in globally hyperuniform states that lack two-component hyperuniformity due to the non-vanishing of inter-correlations (i.e., $\widetilde{S}(0)\neq 0$).

\subsection{Type IV: Single-Component or Two-Component Hyperuniformity with Inter-Correlations}\label{subsec type4}
Sections \ref{subsec com} and \ref{subsec type12} suggest that single-component or two-component hyperuniformity with inter-correlations (Type-IV state) should satisfy either or both of \mbox{Equations (\ref{multi1}) and (\ref{multi2})} with $|D(0)|$ of finite extent.
Therefore, either Equation (\ref{partial zero1}) or Equation (\ref{partial zero2}) is fulfilled at $k_1\rightarrow 0$, yielding
\begin{align}
\lim_{k_1\rightarrow 0}D(k_1)\leq 0
\label{result ineq4}
\end{align}
due to Equation (\ref{ineq3}).
The above relation, similar to Equation (\ref{result ineq3}), indicates that uniform two-component systems become marginal or unstable when single-component or two-component hyperuniformity is preserved while maintaining large-scale correlations between different species.

Plugging Equation (\ref{ineq3}) into Equation (\ref{tilde s by c}), we have
\begin{align}
\widetilde{S}(0)=-\frac{1}{\sqrt{n_{\alpha}n_{\beta}}\,\widetilde{c}(0)}.
\label{multi inter}
\end{align}
The above expression states the following:
there remain inter-correlations unless the magnitude of $\widetilde{c}(0)$ goes to infinity, leading to a lack of global hyperuniformity while maintaining two-component hyperuniformity (Type IV(b)), not to mention single-component hyperuniformity (Type IV(a)).

\subsection{Type V: Multihyperuniformity as the Opposite Limit of Type II}\label{subsec type5}
We consider the following limit of inter-DCF:
\begin{align}
\lim_{k\rightarrow 0}|\widetilde{c}(k)|=\infty,
\label{tilde c inf}
\end{align}
with the finite intra-DCFs, $c(k)$ and $c'(k)$.
In other words, Type-V systems satisfy
\begin{align}
\lim_{k\rightarrow 0}\frac{\mathcal{C}_{\alpha}(k)\,\mathcal{C}_{\beta}(k)}{n_{\alpha}n_{\beta}\widetilde{c}(k)^2}
=0.
\label{limit zero}
\end{align}
It follows from Equations (\ref{tilde c inf}) and (\ref{limit zero}) that the stability indicator defined by\linebreak Equation~(\ref{def D}) becomes
\begin{align}
\lim_{k\rightarrow 0}D(k)=-\infty,
\label{divergence D-}
\end{align}
yielding
\begin{align}
\lim_{k\rightarrow 0}S(k)=\lim_{k\rightarrow 0}S'(k)=\lim_{k\rightarrow 0}\widetilde{S}(k)=0
\label{multi type5}
\end{align}
as seen from the definitions in Equations (\ref{s by c}) to (\ref{tilde s by c}).
Combining Equations (\ref{rep s total}) and (\ref{multi type5}), we have
\begin{align}
\lim_{k\rightarrow 0}S_+^{\mathrm{tot}}(k)=\lim_{k\rightarrow 0}S_-^{\mathrm{tot}}(k)=0:
\label{multi type5(2)}
\end{align}
Type-V states are multihyperuniform.

A comparison between Equations (\ref{divergence D}) and (\ref{divergence D-}) reveals that Type V covers the opposite limit to Type II.
Remembering \mbox{Equations (\ref{result ineq4}) and (\ref{multi inter})}, we also see that Type V includes a limiting state of Type IV(b). 

\subsection{Assessment of Thermodynamic Stability Using Equations (\ref{def variance}) and (\ref{def chi})}\label{subsec therm}

First, we consider Equation (\ref{def variance}) to assess the thermodynamic stability of the hyperuniform states discussed in Sections \ref{subsec type12}--\ref{subsec type5}.
It is straightforward to confirm that the present states, excluding the Type-IV state, satisfy Equation (\ref{def variance}) due to the\linebreak following~reasons: 
\begin{itemize}
\item Type I: Equation (\ref{s'2}) ensures that $S_+^{\mathrm{tot}}(0)>0$.
\item Types II and V: multihyperuniformity necessarily leads to $S_+^{\mathrm{tot}}(0)=0$.
\item Type III(a): G1-hyperuniformity is defined by $S_+^{\mathrm{tot}}(0)=0$.
\item Type III(b): G2-hyperuniformity yields $S_+^{\mathrm{tot}}(0)>0$, as proven in Equation (\ref{result ineq2}) for $\sigma=-$.
\end{itemize}
The exceptional case is Type IV, where Equation (\ref{def variance}) additionally requires $\widetilde{S}(0)>0$ for TC-hyperuniformity.

To meet the second condition (\ref{def chi}) for $\chi_T$, on the other hand, not only Type IV but also Type III(a) must be considered separately.
Let us first examine the other states:
\begin{itemize}
\item Types I, II, and V: all types provide $\chi_T=0$ because of $|D(0)|\rightarrow\infty$. 
\item Type III(b): G2-hyperuniformity amounts to $\chi_T\rightarrow\infty$ because the denominator in Equation (\ref{def chi}) goes to zero by definition of this state.
\end{itemize}
Regarding the exception of Type IV, the relation (\ref{def chi}) additionally requires \mbox{$\widetilde{S}(0)>0$} in Equation (\ref{multi inter}), since the denominator in Equation (\ref{def chi}) equals $-2\sqrt{\phi_{\alpha}\,\phi_{\beta}}\,\widetilde{c}(0)$ when considering TC-hyperuniformity as before.
In Type III(a), the denominator in Equation (\ref{def chi}) reduces to $-4\sqrt{\phi_{\alpha}\,\phi_{\beta}}\,\widetilde{c}(0)$, yielding the additional constraint $\widetilde{c}(0)<0$ due to the relation (\ref{def chi}).
It follows from Equations (\ref{tilde s by c}) and (\ref{result ineq3}) that the condition (\ref{result ineq2}) imposed on G1-hyperuniformity ($\sigma=+$) is reconciled with the constraint $\widetilde{c}(0)<0$ if only $D(0)=0$ (or $\Delta\mathcal{C}(0)=0$);
otherwise, Equation (\ref{rep s -total}) for $\sigma=+$ allows us to verify $S_-^{\mathrm{tot}}(0)<0$ because of $\widetilde{S}(0)>0$, in contradiction to the relations (\ref{result ineq1}) and (\ref{result ineq2}) for $\sigma=+$.

Thus, thermodynamic stability requires no additional conditions for five states except for Types III(a) and IV.
In the exceptional two cases, however, while $S_-^{\mathrm{tot}}(0)\rightarrow \infty$ at $\Delta\mathcal{C}(0)=0$ is imposed on G1-hyperuniformity of Type III(a), $\widetilde{S}(0)>0$ is necessary for TC-hyperuniformity of Type IV.
In other words, G1-hyperuniformity tends to occur accompanied by spinodal decomposition, whereas Type-IV(b) mixtures that violate global-hyperuniformity or multihyperuniformity but preserve TC-hyperuniformity emerge in aggregating systems rather than phase separation (see also Figure \ref{fig heterogeneity}).



\section{Connection with Experimental and Simulation Results}\label{connection}
\subsection{Target Systems for Comparison}\label{subsec target}
Table \ref{table six states} summarizes the characteristics of seven hyperuniform states belonging to the five types (Types I to V).
This table divides the seven hyperuniform states into two major groups depending on whether inter-correlations (ICs) exist or not using $\widetilde{S}(0)=0$ that indicates no macroscopic ICs between two components.
There are three states in the presence of IC where thermodynamic stability requires additional conditions (ACs).
We can define the seven hyperuniform states using the following hyperuniformities:
single-component hyperuniformity (SC), two-component hyperuniformity (TC), global hyperuniformity with $\sigma=+$ (G1), and global hyperuniformity with $\sigma=-$ (G2).
The section numbers represent where this paper discusses the seven states.

\begin{table}[H]
\caption{Classification
table of seven hyperuniform states belonging to Types I--V. The abbreviations are defined as follows (see also the main text or the nomenclature): `ICs' (inter-correlations), `ACs' (additional conditions for thermodynamic stability), `SC' (single-component hyperuniformity), `TC' (two-component hyperuniformity), `G1' (global hyperuniformity with $\sigma=+$), and `G2' (global hyperuniformity with $\sigma=-$).
}
\label{table six states}
\begin{center}
\includegraphics[
width=14cm
]{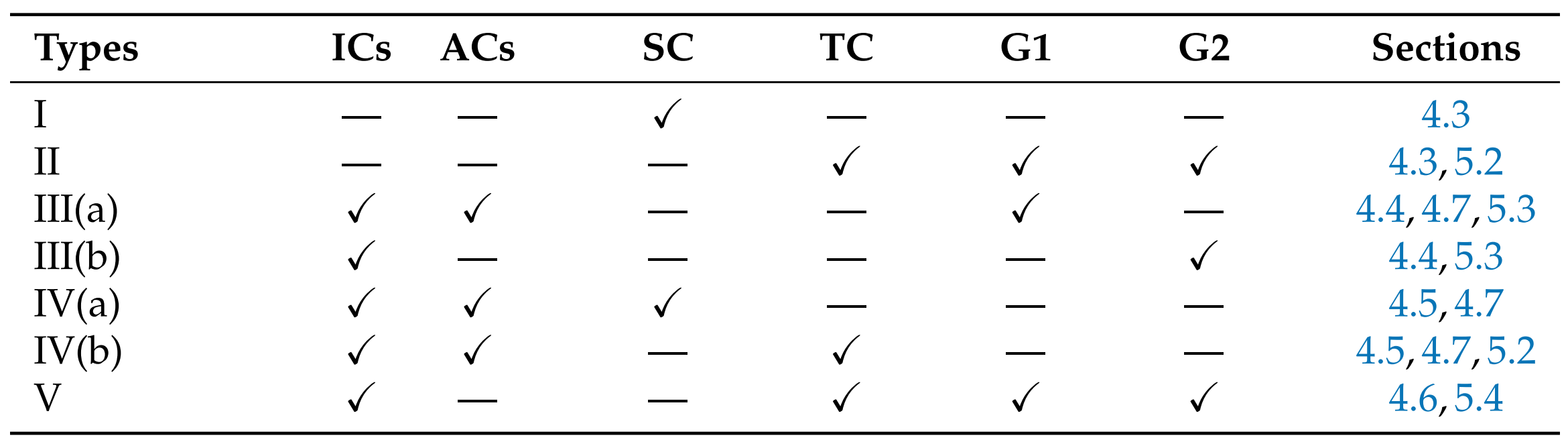}
\end{center}
\end{table}

As illustrated in Figures \ref{fig summary} and \ref{fig heterogeneity}, while Types I and II correspond to homogeneous mixtures due to the lack of correlations between two components, Types III to V exhibit heterogeneity caused by the inter-correlations.
This section focuses on five hyperuniform states of the seven variations, excluding the SC-hyperuniformity of Types I and IV(a), which can be seen as one-component hyperuniform systems immersed in non-hyperuniform media.
In this section, we classify the hyperuniform states observed in the experiments and simulations of multicomponent systems into the five hyperuniform states (Types II, III(a), III(b), IV(b), and (V)) with the help of Table \ref{table six states} as well as Figures \ref{fig summary} and \ref{fig heterogeneity}.

\subsection{A Condition for Multihyperuniformity: Comparison Between Types II and IV(b)}\label{sub 5-1}
Multihyperuniformity is a state in which not only is the entire system hyperuniform (G1 and G2), but each component is also hyperuniform (TC).
To find the conditions for multihyperuniformity, we compare multihyperuniform states (Type II) with a non-multihyperuniform case where each component preserves hyperuniformity while violating global hyperuniformity (Type IV(b)).
Figure \ref{fig summary} and Table \ref{table six states} state that the presence of inter-correlations (i.e., $\widetilde{S}(0)\neq 0$) degrades multihyperuniformity.
In what follows, we briefly review previous experimental and simulation results regarding Types II and IV(b) for assessing the degradation mechanism.

Multihyperuniform Type-II systems found to date include the avian retina \cite{rev1,mul2} and disordered solids such as medium/high-entropy alloys \cite{rev1,mul10}.
The distributions in the photoreceptor cells of five species or atoms in multicomponent alloys have been investigated in experimental and/or Monte Carlo simulation studies \cite{rev1,mul2}, thereby verifying the hyperuniformity of each component as well as the global hyperuniformity.
A previous simulation study has also demonstrated that ternary mixtures of non-additive hard disks can reproduce the main features of avian photoreceptor patterns by adding long-range logarithmic repulsions between particles of the same species \cite{rev1,mul6}.
Non-additive hard disks exhibit demixing under positive non-additivity and local heterogeneous coordination under negative non-additivity \cite{rev1,mul3,mul4,mul5,mul7,mul8,nonad1,nonad2,nonad3,nonad4,nonad5}.
Without Coulomb interactions between unlike particles, Type-II multihyperuniformity emerges in binary non-additive hard-disk plasmas with both positive and negative non-additivities; however, the above ternary mixtures need to be negatively non-additive to satisfy both the multihyperuniformity and local heterogeneous coordination characteristic of avian photoreceptors \cite{rev1,mul6}.

On the other hand, an example of Type IV(b) is a binary mixture of DNA condensate droplets formed by undergoing phase separation in each component of DNA nanostars~\cite{rev1,mul9}.
Fluorescent images of the binary DNA droplets have shown that the sum of droplet density provides a non-zero SF in the zero-wavenumber limit despite the appearance of hyperuniformity for each component.
That is, multihyperuniformity disappears because of $\widetilde{S}(0)>0$ while maintaining TC-hyperuniformity; the positive sign of $\widetilde{S}(0)>0$ is consistent with the additional condition for thermodynamic stability (see Section \ref{subsec therm}).
Focusing on the interactions between particles of different species, repulsive short-range interactions of the excluded-volume type exist in the above previous systems, irrespective of Types II and IV(a), and yet, there is a difference in the formation processes of hyperuniform structures: no inter-correlations are generated by only the static interactions of the excluded-volume type in multihyperuniform Type-II systems, whereas the non-multihyperuniform Type-IV(a) system has a configuration of DNA droplets additionally influenced by hydrodynamic interactions relevant during non-equilibrium phase separation processes \cite{mul9}.
Considering recent findings that hydrodynamic interactions significantly affect the phase-separation morphology of condensates \cite{tanaka1,tanaka2}, it is reasonable to attribute $\widetilde{S}(0)>0$ to the hydrodynamic interactions absent in Type-II systems.

\subsection{Heterogeneous Mixtures only with Global Hyperuniformity: Phase Separation and Complexation in Type-III Systems}\label{sub 5-2}
This section deals with Type-III systems that are globally hyperuniform despite violating the hyperuniformity of each component: either $S_+^{\mathrm{tot}}(0)=0$ or $S_-^{\mathrm{tot}}(0)=0$ holds though $S(0)\neq 0$ and/or $S'(0)\neq 0$ (see Section \ref{subsec type3} for details).
Equation (\ref{rep s total}) implies that the positive intra-SF in total (i.e., $S(0)+S'(0)>0$) is canceled by either the positive or negative value of intra-SF $\widetilde{S}(0)$, representing a macroscopic correlation between different components.
Equation (\ref{rep s -total}) also reveals that G1- and G2-hyperuniformities are incompatible with each other as far as $\widetilde{S}(0)$ is non-zero.
Figure \ref{fig heterogeneity} illustrates that phase separation and complexation are responsible for G1- and G2-hyperuniformities, respectively.

Some simulation studies have demonstrated that G1-hyperuniformity emerges with the lack of multihyperuniformity when forming clusters of the same species in binary mixtures \cite{rev1,mul3,mul4,mul5,mul7,mul8}.
A common feature of these simulations is the introduction of long-range repulsions to compete with short-range interactions that lead to heterogeneity.
The G1-hyperuniform simulation systems include binary mixtures of small and large colloids interacting via repulsive hard-core Yukawa potential \cite{rev1,mul4,mul8} and symmetric mixtures of non-additive hard disks with repulsive Coulombic interactions on a neutralizing background \cite{rev1,mul3,mul5,mul7}; experimentally, however, we had difficulty finding hyperuniform states in bidisperse colloids deposited on a conductive substrate, partly because of particle exchange with bulk dispersion \cite{rev1,mul8}.
Meanwhile, G2-hyperuniformity can be found from simulation results on critical phenomena of the liquid-vapor transition in symmetric electrolytes \cite{ele1}, though no studies explicitly explore G2-hyperuniformity in heterogeneous mixtures.
G2-hyperuniformity corresponds to global charge neutrality in symmetric electrolytes.
Recently, a careful investigation of hyperuniformity in homogeneous symmetric electrolytes was performed, thereby confirming the presence of multihyperuniformity in terms of the density and charge SFs \cite{ele2}.
Yet, previous simulation studies on critical phenomena in symmetric electrolytes have merely suggested the appearance of G2-hyperuniformity without multihyperuniformity: the results roughly indicate that the charge SF disappears (i.e., $S_-^{\mathrm{tot}}(0)=0$) with the divergence of total density SF (i.e., $S_+^{\mathrm{tot}}(0)\rightarrow \infty$) at zero wavenumber \cite{ele1}, which is consistent with Equation (\ref{inf2}).

Let us discuss the former simulation results on G1-hyperuniformity in more detail.
We compare the results of positive additivity with those of negative additivity \cite{mul3,mul5,mul7}.
The simulation results reveal that Type-II multihyperuniformity transforms into Type-III G1-hyperuniformity, irrespective of the sign of non-additivity, by introducing purely Coulombic interactions between unlike particles in addition to between like particles.
The results also show a difference in the charge SF at zero wavenumber (i.e., $S_-^{\mathrm{tot}}(0)$) with the G1-hyperuniformity preserved: we see divergent behavior ($S_-^{\mathrm{tot}}(0)\rightarrow\infty$) for the positive non-additivity, whereas no divergence is indicated for the negative non-additivity despite having a finite non-zero value.

The above features of non-additive hard-disk plasmas are consistent with the following mesoscale structures observed in simulation studies \cite{mul3,mul5,mul7}.
On the one hand, the addition of repulsive interactions with a tendency to demixing due to positive non-additivity facilitates the formation of phase-separated domains while maintaining G1-hyperuniformity, which amounts to a high negative value of $\widetilde{S}(0)$ reflecting a high degree of demixing \cite{rev1,mul3,mul4,mul5,mul7,mul8,nonad1,nonad2,nonad3,nonad4,nonad5}.
Consequently, the charge SF exhibits divergent behavior in agreement with Equation (\ref{rep s -total}).
On the other hand, clustering is induced even in hard-disk plasmas with negative non-additivity, which favors local heterogeneous coordination due to the interplay of long-range Coulomb repulsion and short-range attraction between unlike particles;
however, the preference for local hetero-coordination inhibits the growth of cluster size to a macroscopic scale \cite{rev1,mul3,mul4,mul5,mul7,mul8,nonad1,nonad2,nonad3,nonad4,nonad5}. The suppression of clustering extent predicts that the negative value of $\widetilde{S}(0)$, an indicator of demixing degree, is smaller than that of positive non-additivity, leading to a difference in the charge SF (i.e., $S_-^{\mathrm{tot}}(0)\propto \widetilde{S}(0)$; see also Equation (\ref{rep s -total})), in agreement with the above simulation results \cite{mul3,mul5,mul7}.

Notably, while the former result on the positive additivity is consistent with the additional condition ($S_-^{\mathrm{tot}}(0)\rightarrow\infty$) for thermodynamic stability (see also Section \ref{subsec therm}), our theoretical framework cannot explain the latter result on the negative additivity, partly because the description of local heterogeneous coordination is beyond the scope of this paper.

\subsection{Two-Phase Systems in the Strong Segregation Limits: An Interpretation in Terms of Type-V Multihyperuniformity}\label{sub 5-3}
We first consider the limit of repulsive interactions between unlike particles.
The condition (\ref{tilde c inf}) for Type V then reads
\begin{align}
-\sigma\int d\bm{r}\,\widetilde{c}(\bm{r})=+\infty
\label{mean dcf inf}
\end{align}
with $\sigma=+$, indicating that the strength of short-range repulsive interactions exhibits divergent behavior, though the relation (\ref{tilde s by c}) between $\widetilde{S}(k)$ and $\widetilde{h}(k)$ allows us to rewrite Equation (\ref{multi type5}) for $\widetilde{S}(0)$ to
\begin{align} 
\int d\bm{r}\,\widetilde{h}(\bm{r})=0.
\label{mean tcf zero}
\end{align}
Equations (\ref{mean dcf inf}) and (\ref{mean tcf zero}) show that Type-V systems with repulsive interactions need to meet the following requirements:
while each component is segregated to the highest degree, any species of segregated domains can be found with equal probability regardless of separation $r$ from a reference point, similar to homogeneous mixtures. 
The above requirements are not satisfied by systems undergoing macroscopic phase separation, as shown in the schematic of Type-III(a) systems in Figure \ref{fig heterogeneity}.
This observation suggests that candidate structures with Type-V multihyperuniformity include some morphologies of microphase separation in the strong segregation limit, such as interpenetrated networks of two phases, where unlike particles are excluded from each other in segregated domains.
This theoretical conclusion is consistent with experimental and simulation results on spinodal decomposition in two-phase systems \cite{rev1,two1,two2,two3,two4,two5,two6,two7,two8} and self-assembled structures in soft matter such as block copolymers \cite{rev1,poly1,poly2,poly3}.

Next, we investigate the relevance of an attractive interaction limit for Type-V multihyperuniformity that employs $\sigma=-$ in Equation (\ref{mean dcf inf}), which raises the question of whether Equation (\ref{mean tcf zero}) holds for tightly bound complexation formed due to the strong attractions between unlike particles.
It is noted that the target systems are ternary mixtures because the complex coacervates should be dispersed in a solvent \cite{coa1,coa2}; nevertheless, we would like to continue the discussion by supposing zero density of solute particles outside coacervates.
Accordingly, while the inter-TCF $\widetilde{h}(\bm{r})$ is positive inside a coacervate or between coacervates, it becomes negative between the inside and outside of coacervates, according to the definition of $\widetilde{h}(\bm{r})$ in Equation (\ref{inter}).
We therefore expect to find a condition that satisfies Equation~(\ref{mean tcf zero}), equivalent to the relation in Equation (\ref{multi type5}).
The experimental results on the suspensions of coacervate droplets formed due to the electrostatic complexation of peptides and nucleotides confirm that the spectral density for volumes of coacervate phase goes to zero at zero wavenumber \cite{mul9}.
Since an equal number of cations and anions is ensured inside electrostatic coacervates \cite{coa1,coa2}, the above result of spectral density reads the SFs for cation and anion densities \cite{mul9} as follows: $S(0)=S'(0)=S_+^{\mathrm{tot}}(0)=0$; this implies that $\widetilde{S}(0)=0$ in agreement with the above discussion for Equation (\ref{multi type5}) or Equation (\ref{mean tcf zero}).

Thus, we have demonstrated that hyperuniformity in two-phase systems (or binary mixtures in the strong segregation limit) can be regarded as Type-V multihyperuniformity characterized by Equations (\ref{multi type5}) and (\ref{multi type5(2)}), irrespective of repulsive and attractive interactions (i.e., $\sigma=+$ and $\sigma=-$).

\section{Conclusions}\label{sec6}
A systematic understanding of the diverse hyperuniform states in multicomponent systems was lacking \cite{rev1,rev2,rev3,frusawa1,two1,two2,two3,two4,two5,two6,two7,two8,poly1,poly2,poly3,ele1,ele2,mul1,mul2,mul3,mul4,mul5,mul6,mul7,mul8,mul9,mul10,non1,non2,non3,non4,non5}.
Therefore, this paper has provided a theoretical basis for investigating the underlying mechanisms of various heterogeneous states in binary mixtures.
We have highlighted the novel finding that hyperuniformity exists alongside the divergence of denominator $D(0)$, as illustrated in Figure \ref{fig summary}.
Specifically, we theoretically demonstrate that hyperuniformity of each component can appear when the numerator is zero (see \mbox{Equations (\ref{partial zero1}), (\ref{partial zero2}) and (\ref{zero read3})}), leading to heterogeneity due to the instability of homogeneous systems, as explained in Sections \ref{subsec type3} and \ref{subsec type4} (see also Equations (\ref{result ineq3}) and (\ref{result ineq4})).
This theoretical framework allows us to organize various hyperuniform states observed in experimental and simulation studies \cite{rev1,two1,two2,two3,two4,two5,two6,two7,two8,poly1,poly2,poly3,ele1,ele2,mul1,mul2,mul3,mul4,mul5,mul6,mul7,mul8,mul9,mul10}.
It turns out that hyperuniform binary mixtures can be classified into five types consisting of seven kinds of states.
Figure \ref{fig summary} and Table \ref{table six states} summarize our results.

Our theoretical findings help us find conditions for hyperuniformity in binary mixtures.
There are two key aspects in specifying the hyperuniformity of concern.
The first aspect is whether to aim for hyperuniformity of individual components (i.e., $S(0)=0$ and/or $S'(0)=0$) or to achieve global hyperuniformity while violating SC- or TC-hyperuniformity.
The second aspect involves determining the strategy to design hyperuniform materials that are homogeneous or heterogeneous.

The fabrication strategies vary depending on the type of hyperuniformity targeted.
Our theoretical results offer guidelines for creating a variety of binary hyperuniform mixtures.
For example, we can find various states to maintain global hyperuniformity, focusing on inter-SF $\widetilde{S}(0)$.
Global hyperuniformity never involves multihyperuniformity as long as $\widetilde{S}(0)$ remains finite; G1-hyperuniformity ($\sigma=+$ in \mbox{Equation (\ref{def s total})}) emerges for $\widetilde{S}(0)<0$, whereas G2-hyperuniformity ($\sigma=-$ in \mbox{Equation (\ref{def s total})}) emerges for $\widetilde{S}(0)>0$ (see also Figure \ref{fig heterogeneity}).
Meanwhile, global hyperuniformity is equivalent to multihyperuniformity when $\widetilde{S}(0)$ vanishes.
There are two strategies for creating multihyperuniform states at $\widetilde{S}(0)=0$ that are opposite to each other:
while homogeneous systems require a negligible magnitude of inter-DCF to satisfy Equation (\ref{upper1}), heterogeneous multihyperuniformity is preserved in the large limit of inter-DCF to satisfy Equation (\ref{limit zero}).
In particular, our theoretical framework enables the explicit recognition of aggregation-induced hyperuniformities (G2-hyperuniformity and multihyperuniformity) of Types III(a) and V \cite{ele1,mul9}. This theoretical understanding provides a novel design guideline for fabricating heterogeneous hyperuniform materials, leveraging attractive interactions between different species.

Some issues remain to be addressed in the future.
First, our study assumed that particle volumes of different species are identical to each other (see Section \ref{subsec com}); therefore, we could apply our theoretical results to assessments of the spectral densities of phase volumes obtained in experimental and simulation studies.
In general, however, the proportionality supposed therein is not valid between the total SFs defined in Equation~(\ref{def s total}) and the corresponding spectral densities due to the difference between particle \mbox{volumes~\cite{rev1,two1,two2,two3,two4,two5,two6,two7,two8,mul7}}.
We have yet to extend the present formulations to cover general cases.
Another future challenge is to develop formulations for mixtures of three or more components.
Particularly in heterogeneous systems, ternary mixtures need to be examined considering solvents.
For instance, it is necessary to explicitly consider the presence of solvents in aggregating systems due to attractive interactions between unlike particles, including the Type-III(b) systems satisfying G2-hyperuniformity (see Figure \ref{fig heterogeneity}) and the attractive Type-V systems preserving multihyperuniformity (see Section \ref{subsec type5}).
Despite the limitations of this research, the theoretical approach presented in this paper paves the way for systematic studies of various hyperuniform mixtures.

\section{Appendix: Derivation of Equations (\ref{s by c})--(\ref{total s by c})}\label{appendix}
The rearrangement of Equations (\ref{oz1}) and (\ref{oz4}) gives
\begin{flalign}
\label{app oz1}
\mathcal{C}_{\alpha}(k)\,h(k)&=c(k)+n_{\beta}\,\widetilde{c}(k)\widetilde{h}(k),
\\
\label{app oz4}
-n_{\alpha}\,\widetilde{c}(k)h(k)&=\widetilde{c}(k)-\mathcal{C}_{\beta}(k)\,\widetilde{h}(k),
\end{flalign}
respectively, using $\mathcal{C}_{\alpha}(k)$ and $\mathcal{C}_{\beta}(k)$ in Equations (\ref{def c alpha}) and (\ref{def c beta}).
The combination of Equations (\ref{app oz1}) and (\ref{app oz4}) yields
\begin{align}
\label{app h}
D(k)\,n_{\alpha}h(k)=n_{\alpha}c(k)\mathcal{C}_{\beta}(k)+n_{\alpha}n_{\beta}\,\widetilde{c}^2(k)
\end{align}
by eliminating $\widetilde{h}(k)$ and
\begin{align}
\label{app tilde h}
0=n_{\alpha}\,c(k)\widetilde{c}(k)+\mathcal{C}_{\alpha}(k)\widetilde{c}(k)-D(k)\,\widetilde{h}(k)
\end{align}
by canceling $h(k)$.
Equations (\ref{app h}) and (\ref{app tilde h}) further read
\begin{flalign}
\label{app h2}
D(k)\,\left\{
1+n_{\alpha}h(k)
\right\}&=\mathcal{C}_{\beta}(k)
\\
\label{app tilde h2}
D(k)\,\sqrt{n_{\alpha}n_{\beta}}\,\widetilde{h}(k)&=\sqrt{n_{\alpha}n_{\beta}}\,\widetilde{c}(k),
\end{flalign}
respectively.
It follows from Equations (\ref{app h2}) and (\ref{app tilde h2}) that $S(k)$ and $\widetilde{S}(k)$, defined by \mbox{Equations (\ref{def s}) and (\ref{def tilde s})}, are expressed as Equations (\ref{s by c}) and (\ref{tilde s by c}), respectively.
We can derive Equation (\ref{s' by c}) similarly.
Inserting Equations (\ref{s by c})--(\ref{tilde s by c}) into Equation (\ref{rep s total}), we obtain Equation (\ref{total s by c}).

\subsection*{Nomenclature}
\renewcommand{\arraystretch}{1.5}
\begin{center}
\begin{tabular}{@{}ll}
SF &\quad structure factor\\
TCF &\quad total correlation function\\
DCF &\quad direct correlation function\\
SC-hyperuniformity &\quad single-component hyperuniformity: $S(0)=0$ or $S'(0)=0$\\
TC-hyperuniformity &\quad two-component hyperuniformity: $S(0)=S'(0)=0$\\
G1-hyperuniformity &\quad global hyperuniformity for the sum density: $S_+^{\mathrm{tot}}(0)=0$\\
G2-hyperuniformity &\quad global hyperuniformity for the density difference: $S_-^{\mathrm{tot}}(0)=0$\\
ICs &\quad inter-correlations representing $\widetilde{S}(0)\neq 0$ for the inter-SF $\widetilde{S}(k)$\\
ACs &\quad additional conditions required by thermodynamic stability
\end{tabular}   
\end{center}

\bibliographystyle{apsrev4-1}

\end{document}